# An investigation of the inverted Hanle effect in highly-doped Si

**Yasunori Aoki, Makoto Kameno, Yuichiro Ando, Eiji Shikoh,**

**Yoshishige Suzuki, Teruya Shinjo, and Masashi Shiraishi***

*Graduate School of Engineering Science, Osaka University, 560-8531 Toyonaka, Japan*

* Corresponding author; Masashi Shiraishi (shiraishi@ee.es.osaka-u.ac.jp)

**Tomoyuki Sasaki and Tohru Oikawa**

*SQ Research Center, TDK Corporation, Nagano 385-8555, Japan*

**Toshio Suzuki**

*AIT, Akita Prefectural R&D Center (ARDC), Akita 010-1623, Japan*

## Abstract

The underlying physics of the inverted Hanle effect appearing in Si was experimentally investigated using a Si spin valve, where spin transport was realized up to room temperature. No inverted-Hanle-related signal was observed in a non-local 4-terminal scheme even the same ferromagnetic electrode was used, whereas the signal was detected in a non-local 3-terminal scheme. Although the origin of the inverted Hanle effect has been thought to be ascribed to interfacial roughness beneath a ferromagnetic electrode, our finding is inconsistent with the conventional interpretation. More importantly, we report that there were two different Hanle signals in a non-local 3-terminal scheme, one of which corresponds to the inverted Hanle signal but the other is ascribed to spin transport. These results strongly suggest that (1) there is room for discussion concerning the origin of the inverted Hanle effect, and (2) achievement of spin transport in a non-local 4-terminal scheme is indispensable for further understanding of spin injection, spin transport and spin coherence in Si. Our new findings provide a new and strong platform for arising discussion of the physical essence of Hanle-related spin phenomena.

Nonmagnetic metals, inorganic semiconductors and graphene are good platforms for spin transport and generation of pure spin current up to room temperature (RT), and observation of the Hanle effect using a non-local 4-terminal (NL-4T) method corroborates spin injection and spin transport in these materials [1-4]. Among these materials, Si is considered to be a pivotal material in semiconductor spintronics, because spin coherence in Si is expected to be good due to lattice inversion symmetry [5]. For proving spin injection and transport, observation of magnetoresistance in the NL-4T method provides the most powerful evidence since this method can exclude spurious signals [6], and observation of the Hanle effect, which is attributed to spin precession around an external magnetic field perpendicular to a plane of spin channels resulting in loss of spin coherence, in the NL-4T method also corroborates spin injection and transport in nonmagnetic channels [1].

Another approach to show spin injection (but not spin transport) in nonmagnetic channels is observation of the Hanle [7,8] and inverted Hanle [9] effects using a non-local 3-terminal (NL-3T) method, which enables investigations of spin coherence without producing spin transport in spin valve devices. The inverted Hanle effect is a reciprocal effect of the Hanle effect, wherein interfacial roughness between Si and a tunneling layer beneath the ferromagnet (FM) induces random alignment of spin directions of accumulated spins due to a local magnetostatic field, and application of an in-plane external magnetic field to the spin channel enables recovery of the spin alignment, resulting in an increase of spin accumulation voltages [9]. Considerable discussion arises on the underlying physics of the Hanle and the inverted Hanle effects appearing in the NL-3T method in Si spintronics, since the number of reports on simultaneous demonstration of spin accumulation and spin transport is strongly limited [10], and most previous studies focused only on the Hanle effects in NL-3T where spin transport was not realized in the samples [11-13]. Only one FM electrode is used for both electric current injection and detection of spin accumulation voltages in the NL-3T method, and hence exclusion of spurious signals such as the anisotropic magnetoresistance, the local Hall effect and additional spurious effects such as spin trap effect [7] may not be completely achieved. In fact, the estimated spin coherence at RT in n-Si with the almost same carrier concentrations is not in good accordance between the NL-3T (140 ps [8]) and the NL-4T (1.3 ns [2]) methods, which remains an open question even though some interpretations have been proposed. Such discrepancy may be due to the fact that the NL-3T method is an indirect method for investigating spin transport since only spin accumulation is realized. The purpose of this paper is to show new findings related with the physical essence of Hanle-related spin phenomena in order to arise fruitful discussion that has not been fully implemented. For achieving the purpose, we investigated the inverted and ordinal Hanle effect measurements appearing in a Si spin valve, where spin transport at RT was successfully demonstrated.

As shown in Fig. 1(a), the present device consists of a P-doped ($\sim 5 \times 10^{19}$cm$^{-3}$) n-type Si channel, with a width and thickness of 21 μm and 80 nm, respectively, equipped with two FM and two nonmagnetic (NM) electrodes, all fabricated onto a silicon-on-insulator (SOI) substrate. The FM electrodes were Fe (13 nm thick) with a MgO tunneling barrier layer (0.8 nm thick), with dimensions of 0.5 × 21 μm$^2$ (FM1) and 2 × 21 μm$^2$ (FM2). The center-to-center distance between the FM electrodes was 1.72 μm. The material of the NM electrodes was Al. Details of the fabrication process are described elsewhere [2]. Spin signals were investigated using a dc electric current, where the negative bias current was set as the spin extraction bias from the Si channel. Figure 1(b) shows a typical non-local MR signal observed in the Si spin device at RT, where apparent hysteresis of the non-local resistance was observed. This indicates that spins are definitely injected and transported in the Si channel, and the simultaneous spin injection and transport at RT are successfully demonstrated in our Si spin valve, which has not been achieved in the previous studies using NL-3T. The inverted Hanle effect was measured by applying an external magnetic field along the $x$-axis [9]. For the Hanle effect measurement, the magnetic field was applied along the $z$-axis. Measurements were typically repeated four times using a four-probe system (ST-500, Janis Research) and a physical property measurement system (PPMS, Quantum Design), and the data obtained were averaged. The intensity of spin signals with Hanle-type spin precession was analyzed using the analytical equation describing the non-local spin resistance with Hanle-type spin precession [14],

$$\frac{V(B)}{I} = \frac{V_0}{I} \exp\left(-\frac{d}{\lambda}\right)(1+\omega^2\tau^2)^{-1/4}$$
$$\times \exp\left[\frac{-d}{\lambda}\left(\sqrt{\frac{\sqrt{1+\omega^2\tau^2}+1}{2}}-1\right)\right] \quad (1)$$
$$\times \cos\left[\frac{\sqrt{(\tan^{-1}(\omega\tau))^2}}{2}+\frac{d}{\lambda}\sqrt{\frac{\sqrt{1+\omega^2\tau^2}-1}{2}}\right],$$

where $d$ is the center-to-center distance between the two ferromagnetic electrodes, $\lambda$ is the spin diffusion length, $\omega = g\mu_B B/\hbar$ is the Lamor frequency, $g$ is the electron $g$-factor ($g = 2$), $\mu_B$ is the Bohr magneton, and $\tau$ is the spin lifetime [15]. In the analyses of the NL-3T measurements, the distance between the ferromagnetic electrodes was set to be zero ($d=0$) in the fitting function.

Figure 2(a) shows the results of the inverted Hanle effect measurements at 8 K in the NL-3T method, and an obvious inverted Hanle signal was detected. The signals saturated when the external in-plane magnetic field exceeded ±2000 Oe. As mentioned above, the increase in the spin accumulation signals is ascribed to the recovery of the spin coherence of spins that are accumulated below FM2. Here, it should be noted that spin accumulation beneath FM2 also occurs in the NL-4T method using the measurement setup shown in Fig.1 (a), because the FM2-NM2 circuit is the detection circuit for the spin accumulation voltages in the NL-4T. One question arises as to whether the same inverted Hanle effect can be detected in the NL-4T when FM2 is used as a spin detector. The same effect should be detected, since the same interfacial roughness between FM2 and the Si will induce the inverted Hanle effect (remember that the direction of the external magnetic field is the same, and the same FM electrode is used for voltage detection [16]). Surprisingly, as shown in Fig. 2(b), only non-local magnetoresistance was observed, and no Lorentzian signal due to the inverted Hanle effect was detected. This finding indicates that the conventional interpretation of the inverted Hanle effect is not fully consistent with this result [9]. An additional investigation supporting the above-mentioned result is as follows: The intensity of the inverted Hanle signal in NL-3T, $\Delta V_{3TIH}$, was defined as shown in Fig. 2(a), and the bias electric current dependence of the, $\Delta V_{3TIH}$ is shown in Fig. 2(c), where the bias electric current was varied from -0.5 mA to -3.0 mA and the $\Delta V_{3TIH}$ is observed to be almost constant. This weak bias current dependence of the $\Delta V_{3TIH}$ is notable. The equation describing the intensity of a nonlocal magnetoresistance signal is [15],

$$\Delta V_{non-local} = \frac{P^2 \lambda}{\sigma S} \cdot I \cdot \exp(-\frac{d}{\lambda}), \qquad (2)$$

where $P$ is the spin polarization, $\sigma$ is the conductivity of Si, $d$ is the distance between FM1 and FM2, and $S$ is the area of the interface. Eq. (2) predicts a linear bias current dependence, which has been experimentally observed in studies using the NL-4T method, where the transport of pure spin current was used for the analyses [17,18]. In fact, the NL-4T spin voltages in our device exhibited the expected dependence as shown in Fig. 2(c). If the inverted Hanle effect is related to the spin accumulation, it is quite natural that the NL-3T inverted Hanle signals exhibits a similar linear dependence; however this is not the case in our experiment. According to the literature [8,12], the amplitude of the NL-3T signals might be controversial. However, it is noteworthy that this controversy means there is still much room for discussion of the physics of the Hanle-related phenomena.

Finally, a quite remarkable result is introduced. It has been reported that the full width at half maximum (FWHM) of the inverted Hanle and the Hanle effect in the NL-3T method are in accordance, but the spin lifetime in n-Si estimated from the value of the FWHM in the NL-3T method is less than 200 ps at low temperatures [9], which is much smaller than the spin lifetime obtained in our previous studies using the NL-4T method (ca. 8 ns [2]). Hence, it is quite important to measure the Hanle effect signal in our device in order to clarify whether the inverted Hanle effect can be ascribed to spin accumulation or not. Figure 3(a) shows the results for the inverted Hanle and the Hanle effects in the NL-3Tmethod; it is noteworthy that whereas two peaks in the Hanle signal were observed (see also the inset in Fig. 3(a); a narrower peak was detected between -200 and 200 Oe, while a broader peak was detected between -4000 and 4000 Oe), only one peak appeared for the inverted Hanle effect. The spin lifetimes corresponding to the Hanle signals in the NL-3T method were estimated to be 4.5 ns and 42 ps, respectively, using Eq. (1). Figure 3(b) shows the NL-4T Hanle signal, where an apparent signal can be seen due to spin precession of the propagating spins in the Si channel, and the spin lifetime was estimated to be 8.8 ns using a revised version of Eq. (1) that takes the electrode width into account. The important point here is that the FWHM of the inverted Hanle signals and the broader signal for the Hanle effect are comparable, suggesting almost the same spin lifetime in both cases. Such agreement of the spin lifetimes estimated from the inverted Hanle and Hanle effects has been reported [9]; however spin transport in Si was not reported. Since our spin valve can exhibit the spin transport, we can directly compare the spin lifetimes estimated using the NL-3T and NL-4T methods in the same device. Notably, only one peak was observed in the NL-4T experiment and the spin lifetime (8.8 ns) is comparable to the longer spin lifetime (4.5 ns) in the NL-3T. This discrepancy can be reasonably explained by spin drift due to application of an electric field to the Si channel in the NL-3T scheme, as investigated in our previous study [19] (Note that our previous study on the comparison of the Hanle signals in NL-3T and NL-4T [10] was implemented using ac lock-in technique, where the positive and the negative bias currents were averaged. Thus, no clear difference in the signals was observed). Meanwhile, the shorter spin lifetime (42 ps) in the NL-3T Hanle measurement, corresponding to the inverted Hanle effect, is independent of spin transport since Hanle signals indicating such a short spin lifetime have never been observed in NL-4T Hanle measurements. Furthermore, the bias current dependence of the signal intensities of the broader Hanle signal was almost constant (see Fig. 3(c)), which is in accordance with that in the inverted Hanle signal but is inconsistent with that of the NL-4T spin signals. Thus, the central result of this study is that our finding is inconsistent with the conventional interpretation of the Inverted Hanle effect, and that further investigation is required to correctly interpret the Hanle signals.

In summary, we conducted a systematic study, using the NL-3T and NL-4T methods, in order to clarify the underlying physics of the inverted Hanle effect in Si spin valves that exhibited simultaneous spin injection and transport at RT. We found that no inverted-Hanle-related signal was observed in the NL-4T methods whereas inverted Hanle signals were observed in the NL-3T method. This is inconsistent with the conventional interpretation of the origin of the inverted Hanle effect. More importantly, we found that there were two different peaks in the NL-3T Hanle signals, one with a width comparable to that of the inverted Hanle effect, and another narrower peak that is similar in width to the Hanle signal observed in the NL-4T geometry. While the narrower NL-3T Hanle signal is reasonably well understood, the origin of the signal related to the inverted Hanle signal is unclear since the estimated spin lifetime is too short. Our experimental findings are inconsistent with the conventional concept of interfacial roughness as the origin of the inverted Hanle effect and strongly suggests that achievement of spin transport in NL-4T is indispensable for further understanding of spin injection, spin transport and spin coherence in Si because much closer attention needs to be paid to the interpretation of the inverted Hanle and Hanle effects.

**References**

1. F. J. Jedema, H. B. Heersche, A. T. Filip, J. J. A. Baselmans, and B. J. van Wees, Nature **416**, 713 (2002).

2. T. Suzuki, T.Sasaki, T.Oikawa, M.Shiraishi, Y.Suzuki, and K.Noguchi, Appl Phys. Express. **4**, 023003 (2011).

3. T. Uemura, T. Akiho, M. Harada, K. Matsuda and M. Yamamoto, Appl. Phys. Lett. **99**, 082108 (2011).

4. N. Tombros, C Jozsa, M. Popinciuc, H.T. Jonkman and B.J. van Wees, Nature **448**, 571 (2007).

5. B. Huang, D.J. Monsma and I. Appelbaum, Phys. Rev. Lett. 99, 177209 (2007).

6. M. Ohishi, M. Shiraishi, R. Nouchi, T. Nozaki, T. Shinjo and Y. Suzuki, Jpn. J. Appl. Phys. 45, L605 (2007).

7. M. Tran, H. Jaffrès, C. Deranlot, J.-M. George, A. Fert, A. Miard, and A. Lemaître, Phys. Rev. Lett. **102**, 036601 (2009).

8. S. J. Dash, S. Sharma, R. S. Patel, M. P. de Jong, and R. Jansen, Nature **462**, 491 (2009).

9. S. P. Dash, S. Sharma, J. C. Le Breton, J. Peiro, H. Jaffres, J. –M. George, A. Lemaitre, and R.Jansen, Phys. Rev. B. **84**, 054410 (2011).

10. T. Sasaki, T. Oikawa, M. Shiraishi, Y. Suzuki and K. Noguchi, Appl. Phys. Lett. 98, 012508 (2011).

11. C. H. Li, O. M. J. van't Erve, and B. T. Jonker, Nature Communications **2**, 245 (2011).

12. K. R. Jeon, B. C. Min, I. J. Shin, C. Y. Park, H. S. Lee, Y. H. Jo, and S. C. Shin, Appl. Phys Lett. **98**, 262102 (2011).

13. N.W. Gray and A. Tiwari, Appl. Phys. Lett. **98**, 102112 (2011).

14. T. Sasaki, T. Oikawa, T. Suzuki, M. Shiraishi, Y. Suzuki, and K. Noguchi: IEEE Trans.

Magn. **46**, 1436 (2010).

15. S. Takahashi and S. Maekawa, Phys. Rev. B. **67**, 052409 (2003).

16. We implemented a simple calculation of a spin accumulation voltage in NL-4T and NL-3T based on a phenomenological theoretical model by A. Fert and H. Jaffres (Phys. Rev. B 2001), and found that the Inverted Hanle signal can be observed in NL-4T if the Inverted Hanle effect is attributed to spin accumulation.

17. M. Shiraishi, Y. Honda, E. Shikoh, Y. Suzuki, T. Shinjo, T. Sasaki, T. Oikawa and T. Suzuki, Phys. Rev. B **83,** 241204 (2011).

18. M. Shiraishi, M. Ohishi, R. Nouchi, T. Nozaki, T. Shinjo, and Y. Suzuki, Adv. Func. Mater. **19**, 3711 (2009).

19. M. Kameno, E. Shikoh, T. Oikawa, T.Sasaki, T. Suzuki, Y.Suzuki and M.Shiraishi, in submission (cond-mat arXiv; 1110.4187).

# Figure captions

**Figure 1**

(a) Typical sample structures and measurement geometries in NL-3T (upper panel), and NL-4T (lower panel) methods. An external magnetic field was applied along the $z$-axis for the Hanle and along the $x$-axis for the inverted Hanle measurements.

(b) Typical non-local magnetoresistance signal observed at room temperature, where the bias electric current was set to -3.0 mA. The red and black solid lines show spin accumulation signals during forward and reverse sweeping of the external magnetic field, respectively.

**Figure 2**

(a) Inverted Hanle signal in the NL-3T method at 8 K, where the bias electric current was set to be -3.0 mA. The signal intensity is defined as $\Delta V_{3TIH}$, which is explained in detail in the main text.

(b) Non-local magnetoresistance in the NL-4T method at 8 K, where the bias electric current was set to be -3.0 mA. The inset shows the expanded data around zero magnetic field, where no inverted-Hanle-related signal is observed.

(c) Bias electric current dependences of $\Delta V_{3TIH}$ and $\Delta V_{NL-MR}$ at 8 K, where the bias electric currents were changed from -3.0 mA to -0.5 mA in 0.5 mA steps.

**Figure 3**

(a) Inverted Hanle signal (red open circles) and Hanle signal (green open circles) in the NL-3T method at 8 K, where the bias electric current was set to be -3.0 mA. The inset shows an expanded graph around zero magnetic field. The black solid lines are theoretical fitting lines, and the black dashed line is as a guide to the eye in order to show the narrower peak that is also shown in the inset.

(b) Hanle signal measured using the NL-4T method at 8 K, where the bias electric current was set to be -3.0 mA. Note that only one peak was observed in this measurement.

(c) Bias electric current dependence of the broader Hanle signal appeared in the NL-3T measurement ($\tau$~42 ps) at 8 K, where the bias electric currents were changed from -3.0 mA to -0.5 mA in 0.5 mA steps.

**Figures**

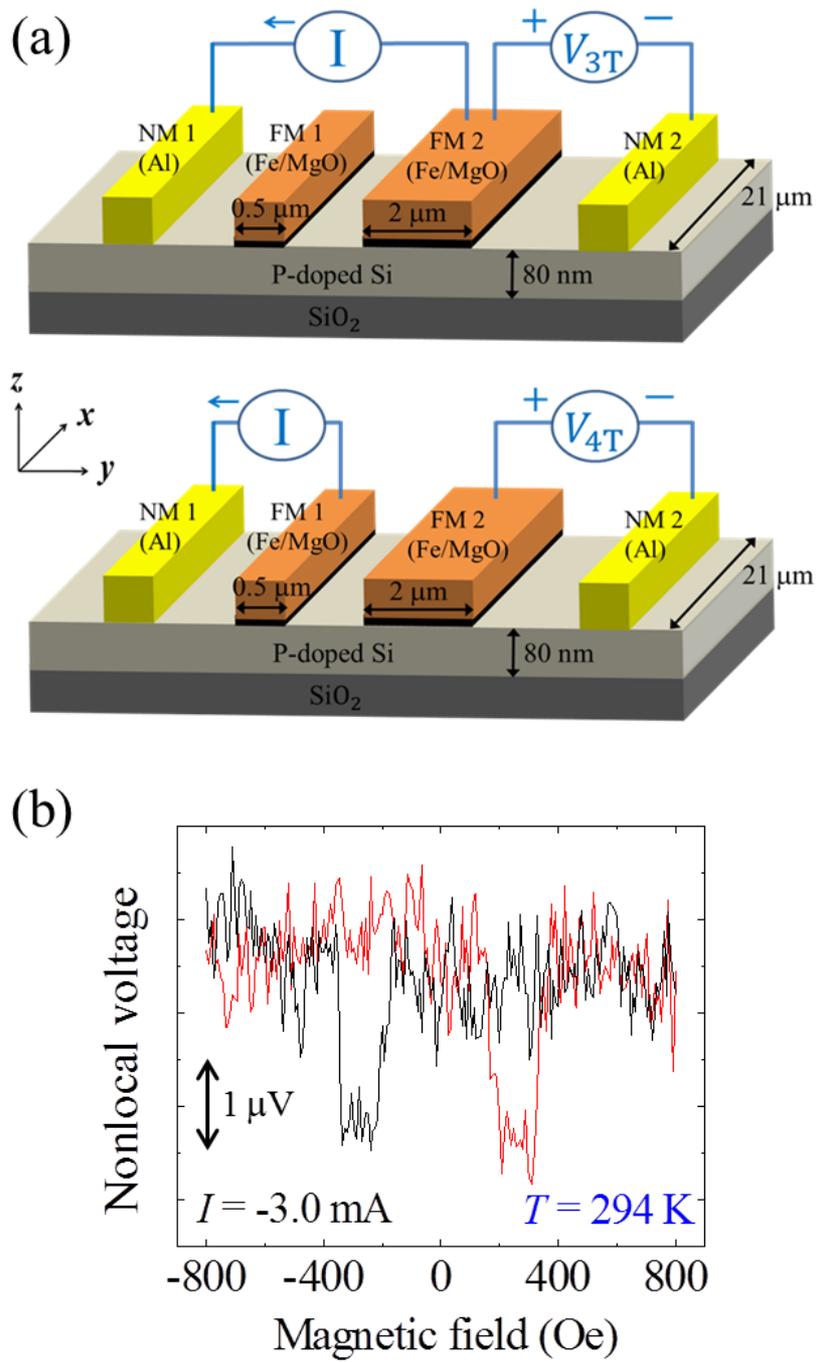

Fig. 1 Y. Aoki et al.

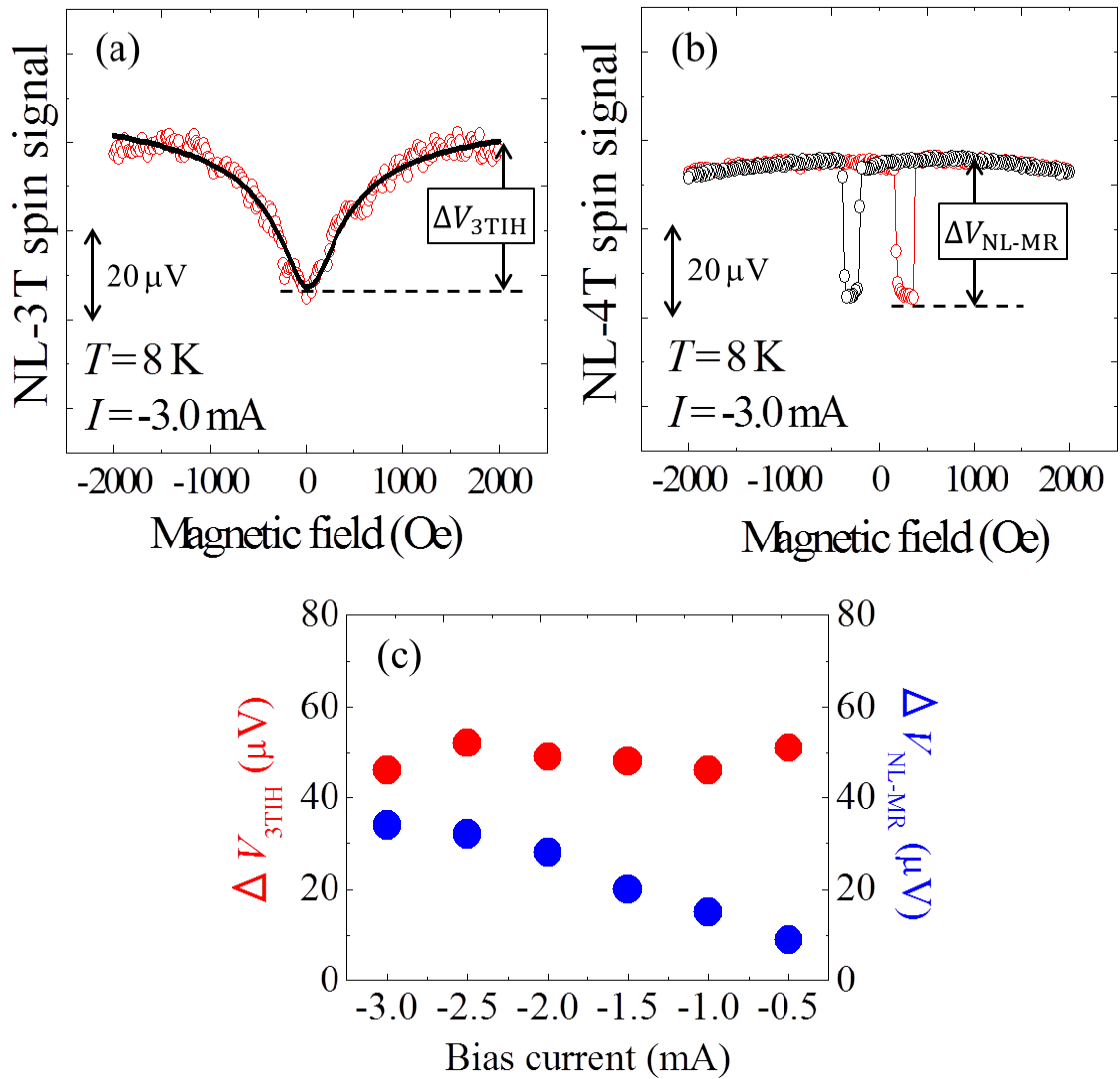

Fig. 2 Y. Aoki et al.

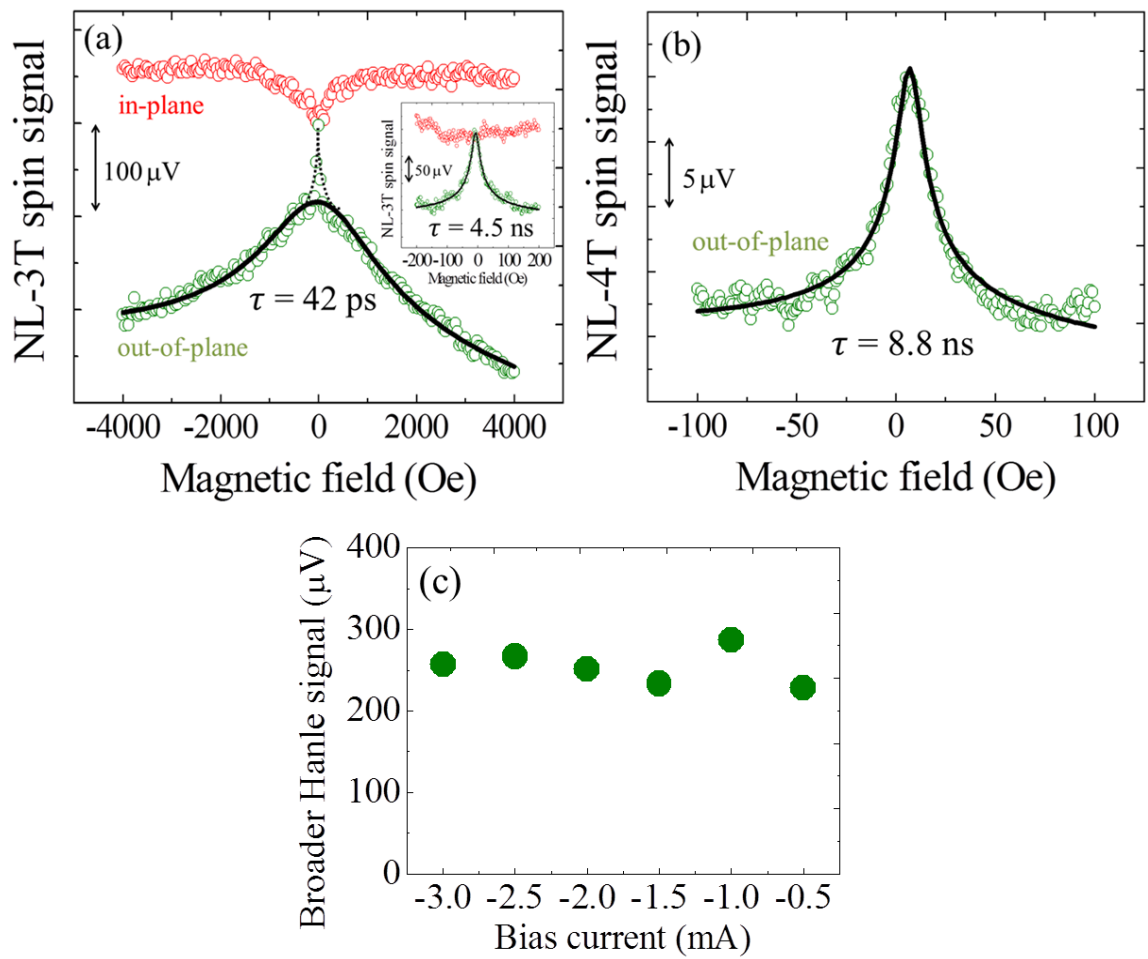

Fig. 3 Y. Aoki et al.